\documentclass[aip, jcp, amsmath,amssymb, reprint, footinbib,floatfix,citeautoscript, longbibliography]{revtex4-1}

\usepackage{graphicx}
\usepackage{dcolumn}
\usepackage{bm}
\usepackage{braket}

\usepackage[dvipsnames]{xcolor}
\usepackage[colorlinks=true,
citecolor=purple,
linkcolor=NavyBlue,
urlcolor=NavyBlue]{hyperref}

\usepackage{enumitem}
\usepackage{makecell}
\usepackage{relsize}
\usepackage{hyperref}

\usepackage{todonotes}

\usepackage{upgreek}

\begin{document}

\author{Anthony J. Dominic III}
\email{Anthony.DominicIII@colorado.edu}
\thanks{These authors contributed equally}
\affiliation{Department of Chemistry, University of Colorado Boulder, Boulder, CO 80309, USA\looseness=-1}

\author{Nicholas L. Cipolla}
\email{ncipolla@trinity.edu}
\thanks{These authors contributed equally}
\affiliation{Department of Chemistry, Trinity University, San Antonio, TX 78212, USA}

\author{William C. Pfalzgraff}
\email{w.pfalzgraff@chatham.edu}
\affiliation{Department of Chemistry, Chatham University, Pittsburgh, PA 15232, USA}

\author{Jeffery A. Jankowski}
\email{jajankowski@noctrl.edu}
\affiliation{Department of Chemistry, North Central College, Naperville, IL 60540, USA}

\author{Rebecca J. Rapf}
\email{rrapf@trinity.edu}
\affiliation{Department of Chemistry, Trinity University, San Antonio, TX 78212, USA}

\author{Andr\'{e}s Montoya-Castillo}
\email{Andres.MontoyaCastillo@colorado.edu}
\affiliation{Department of Chemistry, University of Colorado Boulder, Boulder, CO 80309, USA\looseness=-1} 

\title{A pedagogical tour of the Fourier transform with applications to NMR and IR spectroscopy}

\date{\today}

\begin{abstract}
The Fourier Transform (FT) is a fundamental tool that permeates modern science and technology. While chemistry undergraduates encounter the FT as early as the second year, their courses often only mention it in passing because computers frequently perform it automatically behind the scenes. Although this automation enables students to focus on `the chemistry', students miss out on an opportunity to understand and use one of the most powerful tools in the scientific arsenal capable of revealing how time-dependent signals encode chemical structure. Although many educational resources introduce chemists to the FT, they often require familiarity with sophisticated mathematical and computational concepts. Here, we present a series of three self-contained, Python-based laboratory activities designed for undergraduates to understand the FT and apply it to analyze audio signals, an infrared (IR) spectroscopy interferogram, and a nuclear magnetic resonance (NMR) free induction decay (FID). In these activities, students observe how the FT reveals and quantifies the contribution of each frequency present in a temporal signal and how decay timescales dictate signal broadening. Our activities empower students with the tools to transform their own temporal datasets (e.g., FID) to a frequency spectrum. To ensure accessibility of the activities and lower the barrier to implementation, we utilize Google Colab's open-source, cloud-based platform to run Jupyter notebooks. We also offer a pre-laboratory activity that introduces students to the basics of Python and the Colab platform, and reviews the math and programming skills needed to complete the lab activities. These lab activities help students build a qualitative, quantitative, and practical understanding of the FT. 
\end{abstract}

\maketitle

\section{Introduction}
\label{sec:intro}

\begin{figure}[t!]
    \vspace{-0.25in}
    \centering
    \includegraphics[width=\linewidth]{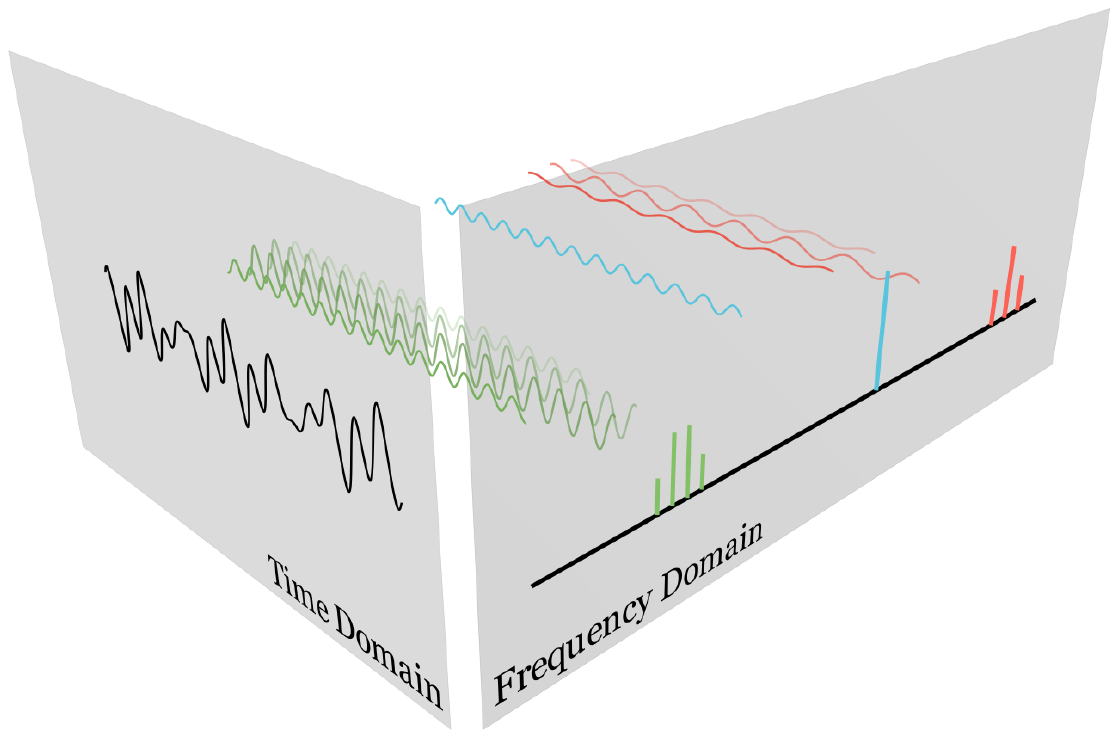}
    \vspace{-0.25in}
    \caption{Illustration of the FT connecting the time and frequency domains. The FT enables one to decouple the sinusoidal frequency contributions to a time-domain signal and captures the `extent' of these contributions in the frequency domain.}
    \label{fig:fig_FT-image}
    \vspace{-0.25in}
\end{figure}

The Fourier Transform (FT) is one of the most powerful and important mathematical tools in the signal processing toolkit. The power of the FT lies in its ability to decompose temporal signals into their constituent frequencies, enabling scientists to analyze and manipulate these components (Fig.~\ref{fig:fig_FT-image}). For these reasons, one can view the FT as a `frequency un-mixing machine'.\cite{3B1B} Moreover, because the FT toolbox provides a general mathematical framework for analyzing signals, the FT is ubiquitously applied across science, including in nuclear magnetic resonance (NMR)\cite{gunther-nmr, borks-book} and infrared (IR)\cite{transform-techniques} spectroscopies, audio processing,\cite{music} machine learning,\cite{brunton-book} magnetic resonance imaging (MRI),\cite{berger-mri, smith-mri} and many more.\cite{discrete-time-signal-processing, transform-techniques, brunton-book} 

In chemistry, we often employ the FT to decompose signals obtained in spectroscopic measurements and assign the influence of chemical identities and environments to peaks in a spectrum. Chemistry undergraduates often first encounter the FT in the context of NMR and IR spectroscopy, as early as sophomore organic chemistry. Yet, despite its fundamental importance and tremendous impact, it is usually treated as a `black box' process in chemistry courses. This is a missed opportunity as all students have had close contact with this sophisticated mathematical tool in their day-to-day lives, and drawing upon these connections presents an exciting learning opportunity to synthesize students' experiences in chemistry and beyond. Specifically, because the FT is an integral part of our senses, it offers an ideal sandbox for students to engage with higher-level mathematics and computer programming \textit{through hearing and vision}. Our pedagogical approach here exploits this realization.

Over the past 50 years, the community has developed various educational materials to elucidate the FT.\cite{glasser-pt1, glasser-pt2, glasser-pt3, graphical-rep} Yet, despite the many articles available, the FT is still typically glossed over in the undergraduate chemistry curriculum. Several reasons may explain this. First, previous works tend to focus on developing either a graphical (qualitative), quantitative, or practical understanding of the FT, instead of combining these as a tool to understand and work with the FT within and beyond chemistry. Second, many existing materials assume a high level of familiarity with advanced math and programming, making them inaccessible to substantive sections of the student population.\cite{LABSIP} To address these issues, we have designed three Python-based activities aimed at third- and fourth-year undergraduate chemistry students enrolled in analytical or physical chemistry courses and centered on building the trifecta of qualitative, quantitative, and practical understanding of the FT. To ensure that these activities can appeal to students at all levels, we have also developed a set of modular pre-laboratory exercises that review trigonometry and calculus concepts and progressively introduce students to Python coding.

We have tested these activities as part of a three-day `Math and Physical Chemistry BootCamp' for first-year physical chemistry graduate students and in two upper-division physical chemistry lab courses, CHEM~4250 at Trinity University and CHM~317L at Chatham University. As written, the activities require two three-hour laboratory sessions. However, we have designed these activities to be modular and also provide a one-hour streamlined version that an instructor could incorporate into a single lecture session (see \texttt{Streamlined.ipynb} in the Supporting Information).

\section{Theoretical Background}
\label{sec:theorybackground}
We designed the lab activities in Sec.~\ref{sec:lab-design} to be accessible to junior and senior chemistry undergraduate students. Specifically, we do not assume that students have any prior experience with the FT, and we have found that students can complete the activities with no prerequisites other than qualitative familiarity with spectroscopy obtained in undergraduate organic chemistry and familiarity with the tools of calculus. Nevertheless, while a detailed theoretical understanding of the FT is not required for students, we provide a brief, pedagogical overview of the theory of the FT~\cite{comp-physics-newman, num-recipes, transform-techniques} and its applications to NMR\cite{gunther-nmr,borks-book} and IR\cite{IR-book} spectroscopy for the benefit of the instructor. These can also be offered to interested students after they complete the activities for further discussion and enrichment. We describe the design and implementation of the lab activities in Sec.~\ref{sec:lab-design}.

\subsection{Fourier Transforms}
\label{ssec:FT-theory}

Although we commonly employ the FT to decompose a time-domain signal, $f(t)$, into its frequency components, $\hat{f}(\omega)$, its applicability is far more general. Fundamentally, the FT connects `conjugate variables', e.g., time and frequency, or position and momentum. Additionally, one can apply the FT to periodic functions that repeat over a finite domain (Fourier series) or to aperiodic functions that decay over an infinite domain (Fourier transform). We illustrate both cases in detail in this section. 

Whether considering a periodic or an aperiodic signal, the FT asks two quantitative questions: Q1) Which frequencies are present in a signal? Q2) How much of each frequency is in the signal? Below, we introduce the background mathematical machinery required to address these questions.

\subsubsection{Fourier Series}
\label{sssec:finite-domain}

We begin by considering a function $f(t)$ defined on an interval $-T \leq t \leq T$ that is periodic with period $p = 2T$, i.e., the same function repeats identically on all intervals $f(t + np) = f(t)$ for any integer $n$. Central to Fourier analysis is the result that one can rewrite $f$ as a \textit{Fourier series}.\cite{brunton-book, boas, fourierseries-georgi} That is, one can decompose $f(t)$ into \textit{even} and \textit{odd} contributions with the \textit{even} parts being a weighted sum of cosines and the \textit{odd} parts being a weighted sum of sines
\begin{equation}
\label{eq:sines-and-cosines}
    f(t) = \sum_{n=0}^\infty a_n \cos\left( \omega_n t \right) +  \sum_{n=1}^\infty b_n \sin\left( \omega_n t \right)
\end{equation}
where $\omega_n = \pi n / T$ is the angular frequency of the sinusoid, and the coefficients $a_n$ and $b_n$ contribute to the \textit{Fourier coefficients} that compose the spectrum. The expansion in Eq.~\eqref{eq:sines-and-cosines} is the Fourier series of $f(t)$. To connect the Fourier series for periodic functions and the FT for aperiodic functions, one can rewrite Eq.~\eqref{eq:sines-and-cosines} into the more compact form
\begin{equation}\label{eq:inverse-F-series}
    f(t) = \sum_{n=-\infty}^\infty \hat{f}_n  e^{i \omega_n t },
\end{equation}
where $i = \sqrt{-1}$ is the complex unit, $e^{ix} = \cos(x) + i \sin(x)$ is Euler's identity for decomposing a complex exponential into sinusoids, and the (possibly complex) Fourier coefficients $\{ \hat{f}_n \}$ are given by the integral
\begin{equation}
\label{eq:fourier-coeffs}
    \hat{f}_n = \frac{1}{T} \int_{-T}^T f(t)  e^{-i \omega_n t } \, {\rm d}t.
\end{equation}
We include this derivation in the Appendix~\ref{ssec:appendix1a} for the interested reader. Here, we decorate $f_n$ with a `hat', $\hat{f}_n$, to denote the FT of the signal $f(t)$. Together, Eqs.~\eqref{eq:fourier-coeffs}~and~\eqref{eq:inverse-F-series} suggest that the FT is invertible, meaning that not only can one compute the spectrum of $f(t)$ using Eq.~\eqref{eq:fourier-coeffs}, one can then subsequently `undo' this transformation using Eq.~\eqref{eq:inverse-F-series} to obtain the original temporal signal.

Qualitatively, one interprets the coefficient $\hat{f}_n$ as the amount of the $n^{\rm th}$ frequency contained in the temporal signal $f(t)$. The mathematical concept that clarifies the meaning of the Fourier integral (Eq.~\eqref{eq:fourier-coeffs}) is the 'orthogonality' of trigonometric functions. Mathematically, if two nonzero functions $f(t) \neq g(t)$ are orthogonal, their product integrates to zero over the interval $-T \leq t \leq T$, i.e., $\int_{-T}^T f(t) g(t) \, {\rm d}t = 0$. However, if this integral evaluates to a nonzero value, then this value tells you `how much' these two functions overlap. With this in mind, we can answer questions Q1 and Q2 from the beginning of this section. 

\begin{figure}[t!]
    \centering
    \vspace{-0.125in}
    \includegraphics[width=\linewidth]{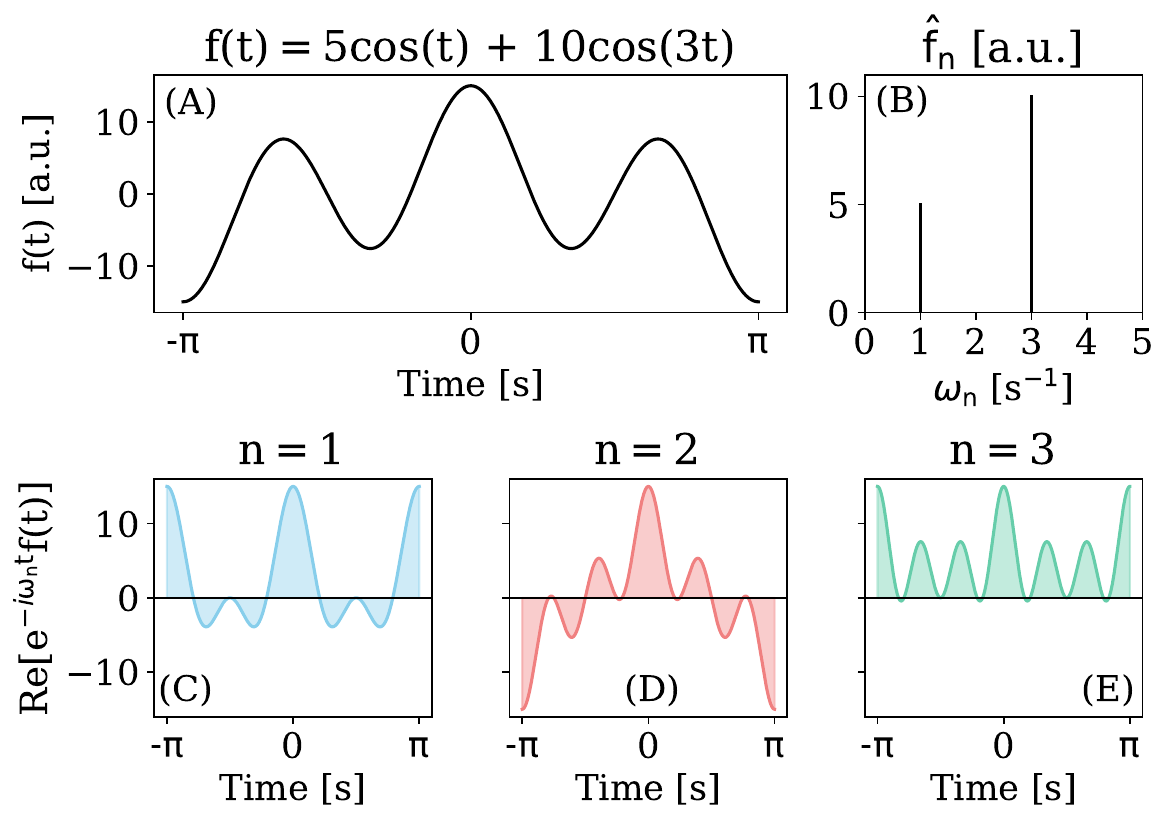}
    \caption{Illustration of the Fourier series (see Sec.~\ref{sssec:finite-domain}). (A) Temporal signal created from the addition of two cosines with two distinct frequencies. (B) Frequency spectrum obtained by computing the Fourier coefficients $\hat{f}_n$ via Eq.~\eqref{eq:fourier-coeffs}. (C)-(E) Demonstrating the orthogonality of the trigonometric functions by considering the real component of the product of the temporal signal $f(t)$ and the complex exponential $e^{-i\omega_n  t}$, where $\omega_n = 2 \pi n / T$ with $T=\pi$~seconds. Note that ${\rm Im}[e^{-i\omega_nt}f(t)] = 0$ in this example. The shaded area corresponds to the integral in Eq.~\eqref{eq:fourier-coeffs}, thus enabling one to visually conclude that $f_1 < f_3$ and that $f_2=0.$ }
    \label{fig:fig_FT-theory}
\end{figure}

In Fig.~\ref{fig:fig_FT-theory}, we demonstrate how the orthogonality concept helps us interpret the Fourier series by considering the periodic temporal signal $f(t) = 5 \cos(t) + 10 \cos(3t)$ on $-\pi \leq t < \pi$ (Fig.~\ref{fig:fig_FT-theory}A). The function $f$ is a combination of waves with frequencies $\omega=1$~s$^{-1}$ and $\omega = 3$~s$^{-1}$. Therefore, we expect to see in $\hat{f}_n$, the FT of $f(t)$, a peak at each frequency represented in $f(t)$ with heights defined by the amplitudes of each sinusoid, as shown in Fig.~\ref{fig:fig_FT-theory}B. To illustrate the origin of each peak in Fig.~\ref{fig:fig_FT-theory}B, we graph the integrand at each of three trial frequencies ($\omega = 1, 2, 3$~s$^{-1}$). For example, in Fig.~\ref{fig:fig_FT-theory}C, we consider $\omega_1 = 1$~s$^{-1}$ and plot the product ${\rm Re}[f(t)e^{-i t}]$. The shaded area under this curve corresponds to how much these two functions overlap and the numerical value of this area is equivalent to the integral in Eq.~\eqref{eq:fourier-coeffs}. As expected, the shaded area is nonzero because there is overlap between our input signal $f(t)$ and $e^{-i t} = \cos(t) - i \sin(t)$ since they both contain $\cos(t)$, and thus we see a peak at $f_1$ in Fig.~\ref{fig:fig_FT-theory}. For similar reasons, we would expect that the area under ${\rm Re}[f(t)e^{-3i t}]$ would be nonzero (Fig.~\ref{fig:fig_FT-theory}E), since both ${\rm Re}[f(t)e^{-3i t}]$ contain $\cos(3t)$. The peak at $f_3$ is larger than $f_1$ because the amplitude corresponding to $\omega_3$ is greater than the amplitude corresponding to $\omega_1$ and thus there is `more of' $\omega_3$ than $\omega_1$ in the input signal $f(t)$. Unlike both of these cases, in Fig.~\ref{fig:fig_FT-theory}D we observe that the area under ${\rm Re}[f(t)e^{-2it}]$ is zero. This is because $f(t)$ does not contain any components with a frequency of $\omega_2 = 2$~s$^{-1}$ and therefore the $e^{-2it}$ and $f(t)$ are orthogonal. Thus, this example shows that the collection of Fourier coefficients $\{\hat{f}_n\}$ are the weights of the specific frequency contributions, and the plot of $\hat{f}_n$ vs.~$\omega_n$ is the frequency spectrum.

\subsubsection{The Fourier Transform}
\label{sssec:infinite-domain}

In the above section, the underlying assumption is that the signal $f(t)$ is periodic with period $2T$, i.e., the same function repeats identically on all intervals $f(t + 2T) = f(t)$. Here we show how to extend this approach to time-series data that are not indefinitely periodic, like the free induction decay measurements obtained in a proton NMR experiment. The trick is to view the signal as having an infinite period length. By applying the limit definition of the integral from calculus\cite{stewart, pons} in conjunction with the limit as $T \rightarrow \infty$ (see Appendix~\ref{ssec:appendix1b}) one can manipulate Eq.~\eqref{eq:fourier-coeffs} into
\begin{equation}
\label{eq:FT}
    \hat{f}(\omega) = \int_{-\infty}^\infty f(t)e^{-i\omega t} \,{\rm d}t,
\end{equation}
the frequency spectrum of our temporal, aperiodic signal. As we mentioned in Sec.~\ref{sssec:finite-domain}, we use the `hat', $\hat{f}$, to denote the FT of the signal $f$. Similar to the discussion in Sec.~\ref{sssec:finite-domain}, one can also `undo' Eq.~\ref{eq:FT} and return back to the original signal $f(t)$ using the \textit{inverse FT}\footnote{For completeness, we note that the above integrals exist provided that $\lim_{t \rightarrow \pm \infty}f(t) = 0$ and $\lim_{\omega \rightarrow \pm \infty}\hat{f}(\omega) = 0$, respectively, and $f$ and $\hat{f}$ are absolutely integrable (i.e., $\int_{-\infty}^\infty |f(t)| \, {\rm d}t < \infty $ and $\int_{-\infty}^\infty |\hat{f}(\omega)| \, {\rm d}\omega < \infty $). Generally, spectroscopic measurements (e.g., FIDs, interferograms) meet these requirements.}
\begin{equation}
\label{eq:ift}
    f(t) = \frac{1}{2\pi} \int_{-\infty}^\infty \hat{f}(\omega) e^{i\omega t} \, {\rm d}\omega.
\end{equation}

\subsubsection{Practical implementation of the FT}
\label{sssec:FT-implement}

In practice, the measured time-domain signal $f(t)$ is sampled over finite, nonnegative values of time $0 \leq t < \infty$ with finite sampling intervals $\Delta t$, and one computes the FT $\hat{f}$ at discrete values of 
$\omega$. Thus, we consider the discrete analogue of Eq.~\eqref{eq:FT} termed the \textit{discrete Fourier transform} (DFT). Mathematically, the DFT is given by
\begin{equation}\label{eq:DFT}
    \hat{f}(\omega_n)  = \sum_{k=0}^{N-1} f(t_k) e^{-i \omega_n t_k},
\end{equation}
where $k$ is an integer time index that takes values from $0$ to $N-1$, and $N$ is the number of measurements in the dataset, and $n$ is an integer frequency index.\cite{num-recipes} 

Although one can implement the FT in Eq.~\eqref{eq:DFT} directly, this approach is often inefficient because calculating the DFT directly at each frequency $\omega_n$ requires evaluating the sum in Eq.~\eqref{eq:DFT} over all values of time $t_k$. That is, evaluating $\hat{f}$ for all frequencies $\omega_n$ computationally requires two nested for-loops: one running over $\omega_n$ and the other over $t_k$. Thus, when one has $N$ values of $t_k$ and $N$ values of $\omega_n$, the total number of operations scales as $N^2$, quantifying the cost of the \textit{direct} algorithm. In computer science language, we say that the computational complexity of the DFT algorithm scales quadratically with the size of the time-domain dataset, i.e.,  $\mathcal{O} (N^2)$ scaling.\cite{num-recipes} We illustrate this scaling issue in Sec.~\ref{sssec:activity1d} to motivate the introduction of ``the most important numerical algorithm of our lifetime"\cite{strang-quote}: the fast Fourier transform (FFT),\cite{num-recipes} which reduces the scaling to $\mathcal{O} (N \log_2 N)$,\cite{num-recipes} rendering the FFT as a significantly more efficient and practical algorithm for computing Eq.~\eqref{eq:DFT}. We refer readers interested in the FFT to Ref.~\onlinecite{num-recipes}. 

\subsection{FT-NMR}
\label{ssec:NMR-theory}

NMR spectroscopy is a topic familiar to all undergraduate chemistry students, but the role of the FT in data collection is rarely a point of emphasis. While, in principle, one can perform NMR spectroscopy using a continuous wave mode with a fixed magnetic field and sequentially collecting data at each frequency (frequency-sweep method) or with a fixed frequency and a variable magnetic field (field-sweep method),\cite{borks-book,gunther-nmr} almost all modern NMR spectrometers use the FT mode that we describe below.

For completeness, we review the basics of NMR spectroscopy for spin 1/2 nuclei to illustrate to the interested student \textit{how} NMR offers a means to distinguish chemical environments. NMR spectroscopy exploits interactions between atomic nuclei (e.g., $^1$H and $^{13}$C) and an external magnetic field ${\bf B}$. In the presence of a magnetic field aligned in the $z$-direction with strength $B_0$, the nuclear spins of the chemical sample either align or anti-align with the direction of the field, leading to an energy difference between those aligned and anti-aligned given by $\Delta E = \gamma \hbar B_0$, where $\gamma$ is the gyromagnetic ratio of the $^1$H (or $^{13}$C) nuclei (values found at Ref.~\onlinecite{moments}). There are slightly more low energy, aligned spins than high energy, anti-aligned spins, giving the sample a bulk magnetization (vector) $M$ along the $z$-direction of the external field $B_0$. In FT-NMR, the sample is subjected to a radiofrequency (RF) pulse that perturbs the external magnetic field and creates a small, oscillating magnetic field in a direction perpendicular to $B_0$. The energy of this RF pulse is tuned to the field-dependent frequency, $f_0$, of the atomic nucleus one is targeting (e.g., $^1$H, $^{13}$C), via $f_0 = \gamma B_0 / 2\pi$.\cite{borks-book, gunther-nmr} This perturbation changes the orientation of the nuclear spins in the sample, tipping the magnetization vector M away from the z-axis. After turning off the RF pulse, the nuclear spins \textit{relax} back to their equilibrium configuration under the constant field, $B_0$. As these spins relax, the bulk magnetization $M$ also relaxes back to its equilibrium orientation along the $z$-axis, and the changing $M$ induces a voltage in the NMR detector coil. We call this signal the free induction decay (FID). The FT of the FID results in the familiar frequency domain spectrum that students learn in organic chemistry.

The spectrum we obtain in $^1$H-NMR experiments provides insight into the molecular structure of a sample. After applying the FT to a temporal signal, a typical spectrum shows intensity as a function of frequency $\nu$ in units of Hertz. However, the convention is to convert this frequency in Hertz to a chemical shift ($\delta$) with units of parts-per-million (ppm) which reports on the change in frequency relative to the reference frequency of tetramethyl silane (TMS), $\nu_{\rm TMS}$, and the operational frequency of the spectrometer, $\nu_{\rm spec}$.\cite{borks-book} Mathematically, $\delta = \tfrac{\nu - \nu_{\rm TMS}}{\nu_{\rm spec}} \times 10^6$. Importantly, this conversion yields a field-independent measurement and implies that the chemical shift does not change from instrument to instrument.

Students are likely familiar with using features of a FT-NMR spectrum (e.g., chemical shifts, $J$-couplings, peak integrations), to characterize the structure of a molecule. It turns out that all of this same information is encoded into the FID, but in a more cryptic way (see Secs.~\ref{sssec:finite-domain} and \ref{sssec:infinite-domain}). To see this, we recall that the FID signal encodes the precession frequency of each nucleus. Because this precession depends on the nuclei's chemical and electrical environments, the detected signal is a sinusoid composed of frequencies unique to each nucleus's environment. These signals combine to form the FID signal that the spectrometer measures at the detector. In addition, this FID signal decays to zero as the nuclear spins relax and approach equilibrium, with the relaxation being dictated by the chemical environment. The timescale of this relaxation is given by the \textit{spin-spin} relaxation time ($T_2$), which helps determine the height and width of each peak in the spectrum.\cite{borks-book} Because the FID signal is a combination of sinusoids with various frequencies and the signal decays to zero, it satisfies the requirements for the FT, as discussed in Sec.~\ref{ssec:FT-theory}.

\subsection{FT-IR}
\label{ssec:IR-theory}

Fourier transform infrared (FT-IR) spectroscopy offers a complementary strategy for characterizing molecules. A basic IR experiment irradiates a chemical sample with a broad range of infrared (IR) light that, upon absorption, excites various vibrational (stretching and bending) modes of molecules in the sample. The detector then measures the IR frequencies that are not absorbed by the sample and are transmitted freely. Before the development of FT-IR, dispersive IR spectrometers collected absorbance data at each frequency sequentially, an approach that was inefficient and resulted in spectra with a low signal-to-noise (S/N) ratio. However, by the 1960s, computers became more efficient and accessible which enabled scientists to design spectroscopy experiments that exploited FT methods, such as FT-IR.

The critical difference between dispersive IR and FT-IR experiments is the Michelson interferometer. This interferometer allows one to probe a sample with all frequencies of IR light \textit{simultaneously} and obtain an interferogram, a temporal signal that encodes the interference of all light transmitted by the sample. The Michelson interferometer uses a beam splitter to split an incident IR light source into two beams: one reflects off a fixed mirror and the other reflects off a moving mirror. The moving mirror causes a path length difference between the two beams. Thus, the two beams interfere (constructively or destructively) when they recombine at the beam splitter. This light then passes through the sample, and the detector measures the total intensity of the transmitted IR frequencies, yielding the desired interferogram. Similar to the FID in an NMR experiment, the interferogram contains the combined information of all transmitted frequencies in the sample. Thus, one can directly apply the FT to disentangle this information and obtain an IR spectrum showing $\%$T as a function of excitation frequency measured in wavenmubers (cm$^{-1}$). However, this spectrum also contains information about the atmosphere surrounding the sample. To isolate the signal of the chemical sample, one performs an additional FT-IR experiment of the atmosphere in the absence of the sample allowing us to obtain a background-subtracted spectrum. Then, one computes the ratio of the transmittance of the sample $T_{\rm s}$ and the transmittance of the background $T_{\rm b}$, i.e., $\% T= T_{\rm s} / T_{\rm b}$.

\section{Lab Design}
\label{sec:lab-design}

We have created a series of Python-based laboratory activities to guide students toward a qualitative, quantitative, and practical understanding of the FT. A distinguishing feature of our approach is that these activities \textbf{do not} require prior coding knowledge (for the student or the instructor). We provide a thorough introduction to Python that discusses each of the tools used in the lab activities. Additionally, we provide keys to the exercises with the Python scripts (Supporting Information).

We have designed the coding activities as fill-in-the-blank prompts that ask students to replace question marks with the correct code or copy-paste-modify code from a previous section. These activities will empower students to modify their Python scripts and inspire them to explore more aspects of the FT toolkit.

We have adopted Google Colab as the user interface for the Python-based activities because it is free, easily accessible, and compatible with all necessary Python packages. Additionally, because Colab is cloud-based, this eliminates the need for students to locally install software on their computers, reducing barriers to implementation. We provide instructions for setting up Colab notebooks in the prelab activity.

\subsection{Prelab}
\label{ssec:prelab}

Students begin with a four-part prelab assignment (see \texttt{Prelab-Questions.pdf} and \texttt{Prelab.ipynb} in the Supporting Information) designed to prepare students with the math and Python concepts necessary for completing the activities. In Prelab~1.1, students focus on useful trigonometry and calculus concepts, including sinusoids, complex numbers, and integration, and employ these to understand the structure of the analytical form of the FT (Eq.~\eqref{eq:FT}). Prelab~1.2 provides a tutorial for students to set up their Google Colab notebook. In Prelab~1.3, students use Google Colab to practice coding in Python. Finally, in Prelab~1.4 students are introduced to the basics of plotting in Python. We have designed these prelab activities to be modular, so that instructors can easily assign only the relevant material for their students, if, for instance, students have prior experience with Python.

\begin{figure*}[t!]
    \centering
    \vspace{-0.125in}
    \includegraphics[width=\linewidth]{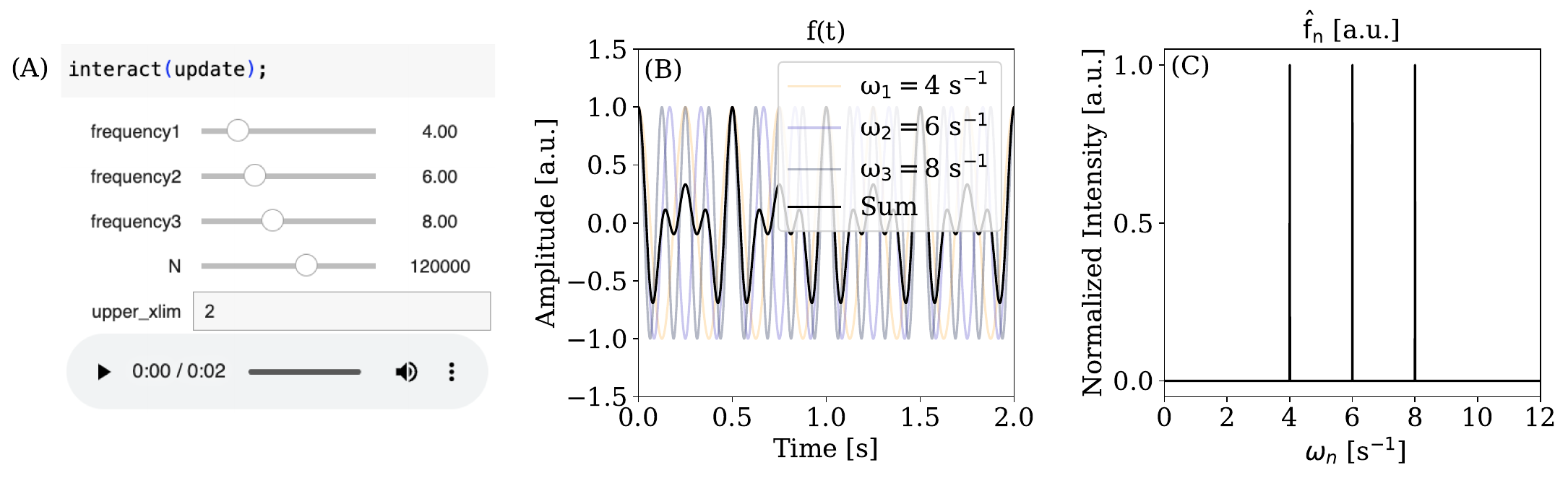}
    \vspace{-0.25in}
    \caption{Sample plots from Activities 1.2-1.3 (See Secs.~\ref{sssec:activity1b}~\&~\ref{sssec:activity1c}). (A) Screenshot from a Jupyter notebook highlighting the interactive portion of Activity~1.2. By adjusting the sliders for each frequency, students can control the frequencies that create the musical chord (the black curve in panel B). Students can also change the duration of the chord by increasing/decreasing the number of sample points $N$ and modify the limits of the $x$-axis in panel (B) by entering different \texttt{upper\_xlim} values. (B) Time-domain signal composed of three sine waves with frequency values of $\omega_1 = 4$~s$^{-1}$, $\omega_2 = 6$~s$^{-1}$, and $\omega_3 = 8$~s$^{-1}$. Students can set each of these values using the sliders in panel (A). (C) The frequency-domain signal obtained via a FT of the time-domain signal in panel (B), showing distinct peaks at each frequency $\{ \omega_1,~\omega_2,~\omega_3 \}$. }
    \label{fig:fig_activity1}
\end{figure*}

\subsection{Activity 1: Applying the FT to audio signals}
\label{ssec:activity1}

Involving multiple senses lowers barriers when acquiring new information.\cite{multisensorylearning} This was the principle in action when the Laser Interferometer Gravitational-Wave Observatory (LIGO) captured the world's attention with an \textit{audible chirp} of the gravitational wave generated by two black holes colliding billions of light years away.\cite{chirp} Because the FT lies at the core of our hearing and vision, we leverage this connection to design Activity~1. In particular, Activity~1 enables students to use the FT to identify the notes within an audio signal of a musical chord.

In Activity~1 (see \texttt{ Activity\_1.ipynb } in the Supporting Information), students begin developing their qualitative understanding of the FT with audio examples by determining the frequencies and amplitudes of the pure tones that contribute to musical chords generated in Python. Students then build a quantitative understanding by focusing on specific computations involving Eq.~\eqref{eq:FT}. This activity guides students to translate math into code, thereby employing a direct implementation of Eq.~\eqref{eq:FT}. This builds student confidence in coding and helps them observe the inefficiency of directly implementing Eq.~\eqref{eq:FT}, as discussed in Sec.~\ref{sssec:FT-implement}. Finally, they explore the practical implementation of the FT using the FFT algorithm, noting the significant increase in computational efficiency.

\subsubsection{Activity 1.1: Visualizing sinusoids}
\label{sssec:activity1a}

\textbf{Learning goal:} \textit{Students will gain familiarity with Python coding, produce high-quality figures in the} \texttt{matplotlib} \textit{Python library, and apply their skills to visualize time domain signals.}

Students begin this activity by utilizing the skills they developed in the Prelab to create sinusoids and explore more complex plotting functions in Python (e.g., defining functions, manipulating and labeling $x$- and $y$-axes, adding grids, etc.). For example, we ask students to consider adding two different sinusoids and plot the result while adding a title, $x$- and $y$-labels, and a figure legend. These skills are essential to the analysis and visualization of datasets in all subsequent exercises.

\subsubsection{Activity 1.2: Listening for qualitative understanding}
\label{sssec:activity1b}

\textbf{Learning goal:} \textit{Students will observe that peak positions in the frequency domain align with the frequencies that compose the sinusoidal temporal signal.}

In Activity 1.2, students focus on qualitatively understanding the FT by analyzing how changes in the time-domain audio signals affect the frequency-domain spectrum. Specifically, students use interactive sliders to control the following parameters: frequency of each wave, length of the signal, and dampening factors. To reinforce sense-based learning, we have designed the activity such that students can listen to the audio generated with these parameters by pressing the `play' button (Fig.~\ref{fig:fig_activity1}A).

\subsubsection{Activity 1.3: Building quantitative understanding}
\label{sssec:activity1c}

\textbf{Learning goal:} \textit{Students will develop familiarity with Python programming by completing and analyzing pre-written code that directly translates the mathematical formulation of the FT in Eq.~\eqref{eq:fourier-coeffs} to code. Students will observe how the FT works numerically by evaluating it at each frequency present in a signal and appreciate qualitatively that each peak in the frequency domain corresponds to a frequency present in the temporal signal. This exercise helps students begin to answer questions Q1 and Q2 from Sec.~\ref{ssec:FT-theory}.}

In Activity 1.3, the students begin exploring the quantitative aspects of the FT, while continuing to develop their qualitative understanding. First, we introduce the DFT (Eq.~\eqref{eq:DFT}) as the discrete analogue of Eq.~\eqref{eq:FT}. Then, the coding exercises guide students to compute the FT of the sinusoids shown in Fig.~\ref{fig:fig_activity1}B. To minimize the complexity of the code while enabling students to understand the numerical components of the FT, students work with code to evaluate the FT (Eq.~\eqref{eq:DFT}) at a single frequency. Specifically, students evaluate the FT at each of the three frequencies, $\omega_1 = 4$~s$^{-1}$, $\omega_1 = 6$~s$^{-1}$, and $\omega_1 = 8$~s$^{-1}$. Importantly, students will find that each computation results in a nonzero value because each of these frequencies are present in the input temporal signal. Students will additionally show that Eq.~\eqref{eq:DFT} is zero when evaluated at $\omega_4 = 10$~s$^{-1}$ since this frequency is not present in the temporal signal. Together, these computations demonstrate the origins of each peak in Fig.~\ref{fig:fig_activity1}C.

\subsubsection{Activity 1.4: Practical advantages of the FFT}
\label{sssec:activity1d}

\textbf{Learning goal:} \textit{Students will utilize the built-in} \texttt{numpy} \textit{Python library to implement the FFT algorithm. Students will compare the speed of the FFT and direct algorithms.}

In Activity 1.4, we introduce students to the practical (i.e., efficient) implementation of the FT while continuing to mine the FT for quantitative insight. We start by quantifying the idea of efficiency. Specifically, students compare the performance of their FT code to that of the FFT algorithm accessible through the \texttt{numpy} library. Students work with a built-in Python package (\texttt{tqdm}) to time their FT and FFT codes for datasets of increasing length. Although the remaining activities utilize the FFT, the intuition gained from the previous exercises is crucial to understanding the conceptual framework of the FT.

\subsection{Activity 2: Application to FT-NMR and FT-IR}
\label{ssec:activity2}

\begin{figure}[b!]
    \centering
    \vspace{-0.125in}
    \includegraphics[width=\linewidth]{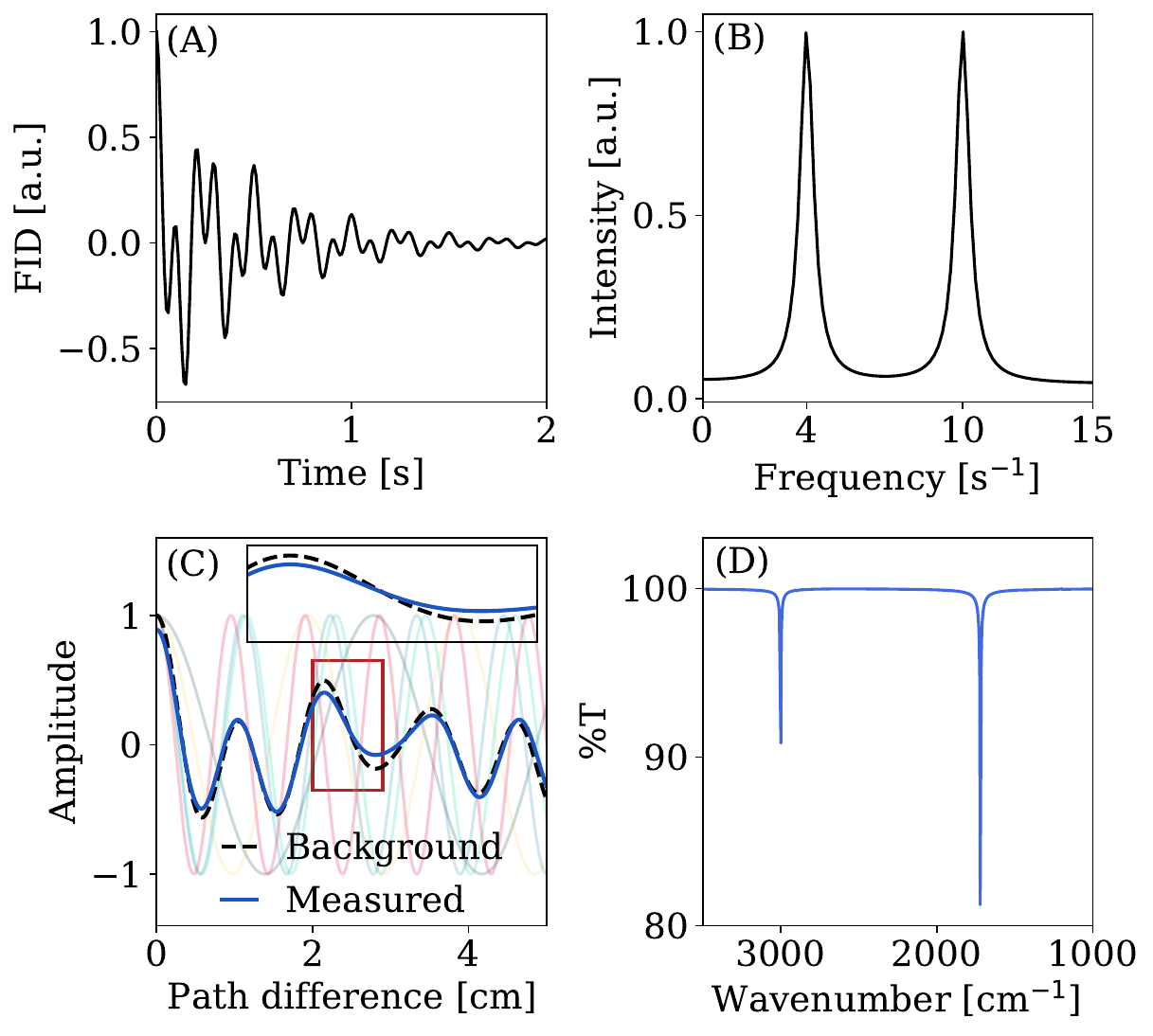}
    \caption{Illustration of the FT in NMR and IR spectroscopy. (A) Model free induction decay (FID) given in Eq.~\eqref{eq:FID} with $\omega_1 = 4$~s$^{-1}$, $\omega_2 = 10$~s$^{-1}$, and $T_2 = 0.5$~s. (B) Frequency spectrum obtained via the FFT of the FID in panel (A) showing distinct peaks centered at both frequencies $\{ \omega_1, \omega_2 \}$ present in the FID. (C) Model interferogram of 3-pentanone constructed from sinuoids with frequencies corresponding to the following wavenumbers: $\tilde{\nu}_1 = 1200$~cm$^{-1}$, $\tilde{\nu}_2 = 1720$~cm$^{-1}$, $\tilde{\nu}_3 = 2900$~cm$^{-1}$, $\tilde{\nu}_4 = 3000$~cm$^{-1}$, $\tilde{\nu}_5 = 3500$~cm$^{-1}$ (translucent curves). The background curve corresponds to a measurement made in the absence of a sample and thus all frequencies are transmitted. The measurement curve corresponds to a measurement made in the presence of a sample where the sample absorbs 20$\%$ of light corresponding to $\tilde{\nu}_2 = 1720$~cm$^{-1}$ and 10$\%$ of light corresponding to $\tilde{\nu}_4 = 3000$~cm$^{-1}$. The inset magnifies the region enclosed in the red box. (D) Background subtracted percent transmittance ($\%T$) spectrum as a function of wavenumbers showing peaks at the frequencies absorbed by the chemical sample.}
    \label{fig:fig_activity2}
\end{figure}

In Activity~2 (see \texttt{Activity\_2.ipynb} in the Supporting Information), students bridge their qualitative and practical understanding of the FT with their experiences in the chemistry classroom. Specifically, in this activity students apply the concepts from Activity~1 to build and analyze FT-NMR and FT-IR datasets. 

\subsubsection{Activity 2.1: Application to NMR spectroscopy}
\label{sssec:activity2-NMR}

\textbf{Learning goal:} \textit{Students will apply the skills from Activity~1 to process a model FT-NMR FID dataset. Students will reinforce their understanding that the position of the peaks in the frequency domain is connected to the sinusoidal frequency components in the FID. Students will observe how increasing/decreasing $T_2$ relaxation time causes peaks to narrow/broaden.}

The students begin by constructing an FID signal consisting of the sum of two sinusoids with different frequencies multiplied by an exponentially decaying function with decay timescale given by $T_2$ introduced in Sec.~\ref{ssec:NMR-theory},
\begin{equation}\label{eq:FID}
    {\rm FID}(t) = e^{\tfrac{-t}{T_2}} [A_1 \cos( \omega_1 t +\phi_1) + A_2\cos(\omega_2 t+\phi_2)],
\end{equation}
where $\{ A_1, A_2 \}$, $\{ \omega_1, \omega_2 \}$, and $\{ \phi_1, \phi_2 \}$ are the amplitudes, angular frequencies, and phases of each sinusoid. We show this FID signal with $A_1 = A_2 = 1$, $\omega_1 = 4$~s$^{-1}$, $\omega_2=10$~s$^{-1}$, and $\phi_1=\phi_2 = 0$ in Fig.~\ref{fig:fig_activity2}A. Because this FID is aperiodic, students will apply the FT as we discussed in Sec.~\ref{sssec:infinite-domain}. The students apply the FFT algorithm to obtain the spectrum in Fig.~\ref{fig:fig_activity2}B. As in Activity~1, students observe that each peak in the spectrum (Fig.~\ref{fig:fig_activity2}B) is centered at one of the frequencies in Eq.~\eqref{eq:FID}. In addition to varying frequencies, students are asked to vary the amplitudes and phases of the sinusoids and the spin-spin relaxation time, $T_2$, and to observe the resultant effects on the spectrum. Students will observe that: (i) changing amplitudes of the FID sinusoids results in a proportional change to the peak heights in the corresponding spectrum; (ii) increasing the phase from $0$ to $2\pi$, the peaks change from slightly negative to completely positive; and (iii) decreasing $T_2$ causes peak widths to grow. 

\subsubsection{Activity 2.2: Application to IR spectroscopy}
\label{sssec:activity2-IR}

\textbf{Learning goal:} \textit{Students will apply the FT to process and analyze a model IR interferogram. This activity will reiterate that peaks in the IR spectrum correspond to the frequencies that transmit through the sample and onto the detector in an IR experiment.}

Here, students construct and analyze a fictitious FT-IR spectrum of 3-pentanone. We say ``fictitious'' because, for simplicity, we reduce the complexity of the light source (experimentally, a Globar) irradiating the sample, to an incident source composed of only 5 frequencies with corresponding wavenumbers given by: 1200 cm$^{-1}$, 1720 cm$^{-1}$, 2900 cm$^{-1}$, 3000 cm$^{-1}$, and 3500 cm$^{-1}$. We choose these values to illustrate two instances where one expects absorption and three instances where no absorption occurs. Students generate the corresponding sinusoids for each wavelength, shown as translucent curves in Fig.~\ref{fig:fig_activity2}C. In the foreground of Fig.~\ref{fig:fig_activity2}C, we illustrate the `background' measurement where one measures all of the incident light at the detector due to the absence of a sample. We then show a `measured' interferogram curve constructed under two assumptions: (i) that our sample absorbs 20\% of the 1720 cm$^{-1}$ light, 10\% of the 3000 cm$^{-1}$ light, and (ii) that all other frequencies of light pass through the sample. Students combine their acquired knowledge of the FT from Sec.~\ref{ssec:FT-theory} and FT-IR theory in Sec.~\ref{ssec:IR-theory} to compute the spectrum of percent transmittance ($\%$T) in Fig.~\ref{fig:fig_activity2}D. Finally, we ask students to consider the differences between this modeled spectrum and an experimental spectrum from Ref.~\onlinecite{3pentanone}, encouraging them to reflect on the coupling of normal modes within a molecule.

\subsection{Activity 3: Processing your own FID}
\label{ssec:activity3}

\begin{figure}[b!]
    \centering
    \vspace{-0.125in}
    \includegraphics[width=\linewidth]{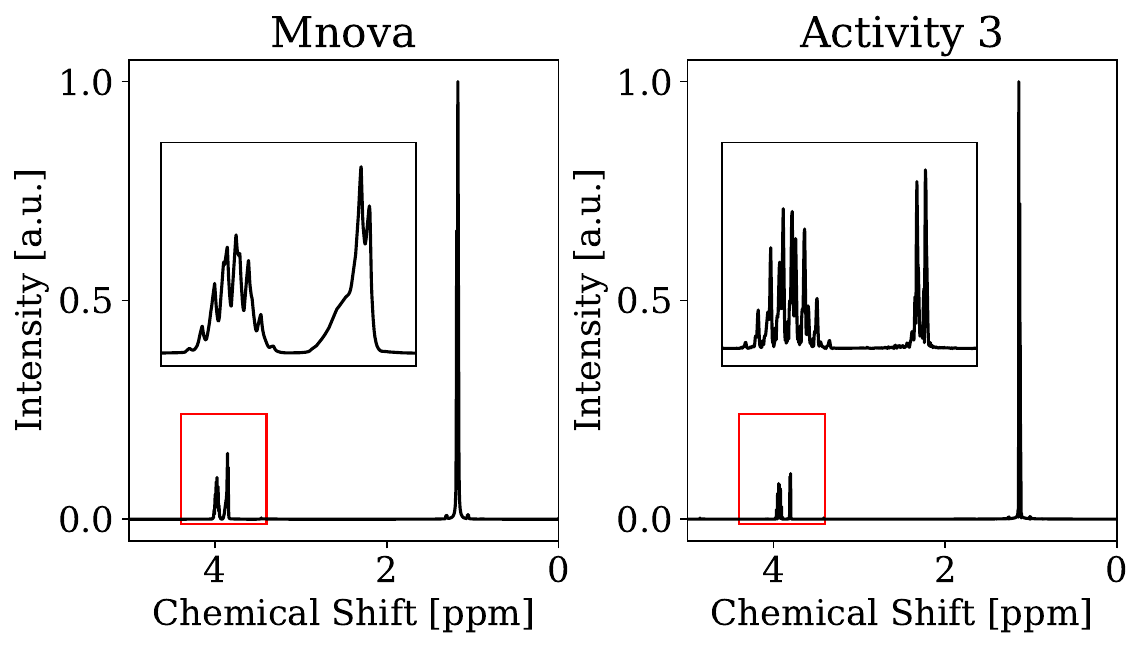}
    \vspace{-0.25in}
    \caption{Comparison of 2-butanone $^1$H-NMR spectra obtained using Mnova (A), a popular post-processing software, and the windowing procedure in activity 3 (B). The insets magnify the region enclosed in the red box.}
    \label{fig:fig_activity3}
    \vspace{-0.125in}
\end{figure}

\textbf{Learning goal:} \textit{Students will synthesize and apply the skills learned in previous activities to transform an FID dataset obtained from an FT-NMR experiment to the familiar FT-NMR spectrum. Students will recall their knowledge of NMR shifts to assign the identity of the molecule. }

In Activity~3 (see \texttt{Activity\_3.ipynb} in the Supporting Information), students synthesize the skills developed in Secs.~\ref{ssec:activity1}~and~\ref{ssec:activity2} to process an FID signal corresponding to an experimental FT-NMR signal for 2-butanone in deuterated chloroform solvent (CDCl$_3$), collected on a Varian 500~MHz NMR spectrometer. We ask students to plot and manipulate the raw FID file obtained from the FT-NMR experiment. Specifically, we first ask students to subtract the average of the FID measurements to ensure that the FID decays to zero at long times, satisfying the FT requirement that ${\rm FID}(t) \rightarrow 0$ as $t \rightarrow \infty$  (Sec.~\ref{ssec:FT-theory}). We then provide code to zero-pad  the end of the temporal signal (i.e., append zeros), and explain that this allows one to adjust the resolution of spectrum as well as ensuring that the length of the FID is a power of 2, maximizing the efficiency of the FFT (Sec.~\ref{ssec:FT-theory}). Finally, we introduce students to the Lanczos and Hanning windowing functions that improve the signal-to-noise (S/N) ratio.\cite{windowing} We invite students to explore the effects of each windowing function as well as the combination or absence of both windowing functions. After post-processing the spectrum, we ask students to compare the processed (windowed) spectrum to the unprocessed spectrum (without windowing). Finally, we provide students with the chemical formula for 2-butanone and ask them to identify the molecule and label the corresponding signals on their spectrum. In Fig.~\ref{fig:fig_activity3}, we display the resulting spectrum that students are expected to generate and compare it to the spectrum we have processed using Mnova,\cite{mnova} a widely used FT-NMR post-processing software. This activity aims to empower students with the tools to process and analyze their own FID datasets.

\subsection{Implementation of this lab activity}

We tested an initial version of these activities at the University of Colorado Boulder as a learning module in a three-day `Math and Physical Chemistry BootCamp' orientation workshop aimed at incoming first-year physical chemistry graduate students (16 students). We employed student and instructor feedback to revise the activities. We have also successfully implemented these activities in two upper-division physical chemistry lab courses, CHEM~4250 at Trinity University (8 students) and CHM~317L at Chatham University (10 students). 

At Trinity, the majority of students had no prior programming experience and were not previously familiar with the mathematics of the FT. At Chatham, students had been doing weekly exercises that included some programming for a few weeks, but for almost all of them, this was their first introduction to the FT. We administered the prelab exercise during class time for students at Trinity, who did not have prior coding experience. All students completed the prelab independently without assistance from the instructor, taking $\sim$30 minutes to complete the math portion (Prelab~1.1) and $\sim$45 minutes to complete the coding portion (Prelab~1.2-1.4). At Chatham and Trinity, students completed the three activities over two successive three-hour lab periods. Each student completed the activities independently. Activity~1 took students $\sim$90 minutes, Activity~2 required $\sim$100 minutes, and Activity~3 required $\sim$120 minutes. 

Students completed all activities during class time, allowing instructors to address any confusion or misconceptions in real-time. Students found the activities to be manageable in difficulty and enjoyed the activities, especially the connection to music in Activity 1. Activity 2 inspired interesting discussions about IR spectroscopy, including experimental instrumentation (output of the Globar), and the coupling between modes observed in the fingerprint region. Most students could correctly identify the unknown chemical at the end of Activity 3. By successfully completing the activities, students gained familiarity and understanding of the FT. Conducting these activities at the beginning of the semester has enabled continued discussions of the FT's role (or lack thereof) in other experiments, including discussing why, for instance, FT is not commonly used for UV-vis spectroscopy.

\section{Suggested Modifications}

We have designed these activities to be modular so that instructors can choose the sections that fit their curriculum or time constraints. For example, if students are already familiar with Python programming, instructors can omit sections that cover the introduction to Python. 

As designed, the three activities take approximately two 3-hour periods and fit well into an upper-division chemistry laboratory course. We recognize that this is a significant time commitment, and encourage instructors to adapt the modules as needed to fit their time constraints. Here, we suggest three specific modifications that reduce the time required for the students to complete these activities. First, instructors can adopt a shorter version of the laboratory by removing either \hyperref[sssec:activity2-NMR]{Activity~2.1} or \hyperref[sssec:activity2-IR]{Activity~2.2}. Removing \hyperref[sssec:activity2-IR]{Activity~2.2}, frames the activities solely with a focus on NMR spectroscopy as an analytical technique. Removing \hyperref[sssec:activity2-NMR]{Activity~2.1} or \hyperref[sssec:activity2-IR]{Activity~2.2} reduces the total activity time by $\sim$50 minutes. 

Second, for instructors with only one $\sim$3 hour lab period available for this activity, we suggest omitting \hyperref[ssec:activity3]{Activity~3}
Activity~3, which reduces the total activity time by $\sim$120 minutes. With this modification, students still investigate the role of the FT in both NMR and IR spectroscopy in Activity~2, but without the additional data processing considerations for ``real'' experimental NMR data given in Activity~3.

Third, instructors in lecture courses are also likely to find the content covered in this activity of interest. For this purpose, we have provided a streamlined version of the activities combining Activities~1 and~2 that is appropriate for a $\sim$1-hour class period. The interested reader can find this streamlined activity (labeled \texttt{Streamlined.ipynb}) in the Supporting Information. In this version, students work through the same conceptual material, but with more guidance in the Python coding and less emphasis on plotting with the \texttt{matplotlib} library. This further reduces any barriers due to lack of familiarity with Python while maintaining the ability to meet the learning objectives of the activity in a condensed format.

For interested instructors or students, there are many possibilities for extending these activities, especially Activity~3. If instructors want to add an experimental component to this activity, students could collect their own experimental NMR data on their molecule of choice, extract the FID, and use that data to conduct Activity~3. This would also lend itself to an exploration of the impact of instrument resolution on the collected FID and resultant spectrum. For example, instructors could expand on Activity~3 by comparing the data collected on a higher resolution instrument (e.g., in Activity~3, we used 500~MHz) to a lower resolution instrument (e.g., a benchtop NMR with 90~MHz). The topics included in these activities serve as an introduction to the Fourier analysis toolkit, on which instructors can expand on in many ways, such as with deeper discussions of the FFT algorithm, zero-padding, and windowing functions. Furthermore, this sets the framework for introducing more complex topics such as aliasing, noise reduction, and phase correction.

\section{Conclusion}

We have designed a series of self-contained, Python-based exercises for undergraduate chemistry students that focus on building a trifecta of qualitative, quantitative, and practical understanding of how the FT converts temporal signals into frequency-domain spectra. By completing these exercises, students can learn and apply the FT to analyze an audio signal, an FT-NMR FID, and an FT-IR interferogram. Our activities tailor opportunities for students to synthesize these skills to transform an FID signal obtained in a real FT-NMR experiment of 2-butanone and apply their chemistry knowledge to assign the chemical identity of the peaks in this spectrum. We have made our approach accessible by utilizing Google Colab Jupyter notebooks to remove the burden that students often face with installing software across different computer systems, difficulties compiling various Python packages, and limited memory/storage for running Python code on their machines. We have ensured that our approach is self-contained by including a set of pre-laboratory exercises to introduce the necessary mathematical and Python programming skills that are useful for completing the activities, and we additionally provide a tutorial for setting up and navigating the Colab notebooks. We created accessible self-contained activities to reduce the barrier to adoption into the undergraduate curriculum. Furthermore, we illustrate the power and versatility of the FT by synthesizing examples in and beyond chemistry and encourage students to apply the FT in their own work and research. Our work provides a modular template that instructors can modify to serve the needs of their analytical, spectroscopy, or physical chemistry courses.

\section{Associated Content}
Our materials are freely available at the following GitHub repository: \url{https://github.com/Montoya-Castillo-Group/FourierTransforms}. Students can run each \texttt{.ipynb} file by manually uploading them to Google Colab.

\section{Acknowledgments}
A.J.D.~acknowledges support from the NIH/CU Molecular Biophysics Program and the NIH Biophysics Training Grant T32 GM145437. N.L.C.~acknowledges the Arnold and Mabel Beckman Foundation Beckman Scholars Program (BSP2022). W.C.P.~acknowledges support from Cottrell Scholar Award CS-CSA-2023-077 sponsored by the Research Corporation for Science Advancement and from Integrated Research-Education Grant KA2022-129526 from the Charles E.~Kaufman Foundation of the Pittsburgh Foundation. R.J.R.~acknowledges financial support from grant CS-CSA-2024-098 from Research Corporation for Science Advancement. A.M.C.~acknowledges the start-up funds from the University of Colorado Boulder. The NMR spectrometer was funded by support from the National Science Foundation (CHE-0957839). The authors thank Dr.~Paolo Suating for help in collecting experimental NMR spectra and Profs.~Meredith Borden, Matthew Pons, and David Schmitz for feedback on the article. We additionally thank Prof.~David Horner for insightful discussions and feedback on the final manuscript.

\section{Appendix}
\label{sec:appendix}

\subsection{Rewriting the Fourier sine and cosine series in terms of complex exponentials}
\label{ssec:appendix1a}

We start with the Fourier `sine and cosine' series of a periodic, smooth function $f$, given by 
\begin{equation}
\label{aeq:sines-and-cosines}
    f(t) = a_0 + \sum_{n=1}^\infty a_n \cos\left( \omega_n t \right) +  \sum_{n=1}^\infty b_n \sin\left( \omega_n t \right),
\end{equation}
where
\begin{equation}
\label{aeq:an-s}
a_n = 
    \begin{cases}
        \dfrac{1}{2T}\displaystyle\int_{-T}^T f(t) \, {\rm d}t, & n =0;\\[1.5em]
        \dfrac{1}{T}\displaystyle\int_{-T}^T f(t) \cos(\omega_n t) \, {\rm d}t, & n \geq 1,
    \end{cases}
\end{equation}
and
\begin{equation}
\label{aeq:bn-s}
    b_n = \frac{1}{T}\int_{-T}^T f(t)\sin(\omega_n t)  \, {\rm d}t,
\end{equation}
for $n \geq 1$. Then, we insert the identities $$\cos(\theta) = \frac{e^{i\theta} + e^{-i\theta}}{2} \,\,\,\text{ and }\,\,\, \sin(\theta) = \frac{e^{i\theta} - e^{-i\theta}}{2i}$$ into Eq.~\eqref{aeq:sines-and-cosines} to obtain
\begin{equation*}
\begin{split}
    f(t) &= \sum_{n=0}^\infty \left[ a_n \cos(\omega_n t) + b_n \sin(\omega_n t) \right] \\
    &= \!\sum_{n=0}^\infty \!\bigg[ 
        \dfrac{a_n}{2} \big( e^{i\omega_n t}\! + e^{-i\omega_n t} \big)  + \dfrac{b_n}{2i} \big( e^{i\omega_n t} - e^{-i\omega_n t} \big) 
    \bigg]\\[0.1em]
    &= \!\sum_{n=0}^\infty \bigg[ \dfrac{1}{2}
         \left( a_n - i b_n\right)e^{i\omega_nt} + \dfrac{1}{2}\left(  a_n + i b_n\right)e^{-i\omega_nt} 
    \bigg]\\[0.1em]
    &= \!\sum_{n=0}^\infty \dfrac{1}{2}
         \left( a_n - i b_n\right)e^{i\omega_nt}\! +\! \!\sum_{n=-1}^{-\infty} \dfrac{1}{2}
         \left( a_{-n} + i b_{-n}\right)e^{i\omega_nt}\\[0.1em]
    &\equiv \frac{1}{T} \sum_{n=-\infty}^\infty \hat{f}_n e^{i \omega_nt},
\end{split}
\end{equation*}
where we have used the fact that $\omega_n = 2 \pi n / T$ re-indexes the sum (i.e., $n \rightarrow -n$) in the penultimate equality, allowing one to define the Fourier coefficients $\{ \hat{f}_n \}$ as
\begin{equation}
\label{aeq:fn-s}
    \hat{f}_n = 
    \begin{cases}
        \dfrac{T}{2}(a_n - i b_n), & \,\,\,n > 0; \\[1em]
        Ta_0, & \,\,\,n=0;\\[1em]
        \dfrac{T}{2}(a_n + i b_n), & \,\,\,n < 0.
    \end{cases}
\end{equation}
Inserting Eqs~\eqref{aeq:an-s} and Eq.~\eqref{aeq:bn-s} into Eq.~\eqref{aeq:fn-s}, one can write the general expression
\begin{equation}
    \hat{f}_n = \int_{-T}^T f(t)e^{-i \omega_n t}\, {\rm d}t.
\end{equation}

\subsection{Fourier series as period length approach infinity}
\label{ssec:appendix1b}

Here, we derive Eqs.~\eqref{eq:FT}~and~\eqref{eq:ift} from Eqs.~\eqref{eq:inverse-F-series}~and~\eqref{eq:fourier-coeffs}. The key is to think of the function $f$ as being periodic with period length approaching infinity, i.e., $T \rightarrow \infty$. Consequently, as $T \rightarrow \infty$ the difference between $\omega_n$ and $\omega_{n+1}$ becomes infinitesimally small. Thus, one can invoke the limit definition of the integral to connect the summation over the discrete index $n$ in the Fourier series to an integral over a continuous variable in the FT.

To start, we introduce a new variable to keep track of the difference between consecutive frequencies, $\Delta = \omega_{n+1} - \omega_n = 2\pi / T$. Then, we insert this into Eq.~\eqref{eq:fourier-coeffs} and obtain
\begin{equation*}
    f(t) = \frac{1}{2 \pi } \!\sum_{n=-\infty}^\infty \!\frac{2\pi}{T} \hat{f}_n \, e^{i \omega_n t} = \!\frac{1}{2\pi } \!\sum_{n=-\infty}^\infty  \!\!\Delta \,  \hat{f}_n \,  e^{i \omega_n t}.
\end{equation*}
As the period length of $f$ becomes infinite, $\Delta = \omega_{n+1} - \omega_n \rightarrow 0$ and thus $\omega_n$ becomes a continuous variable, $\omega$. It also follows that $\hat{f}_n \rightarrow \hat{f}(\omega)$, $\Delta \rightarrow {\rm d}\omega$, and $\sum_n \Delta \rightarrow \int {\rm d}\omega$. Mathematically,
\begin{equation}
    f(t) \!\equiv \frac{1}{2\pi}\int_{-\infty}^\infty \hat{f}(\omega)e^{i\omega t}\, {\rm d}\omega \!= \!\lim_{\Delta \rightarrow 0} \sum_{n=-\infty}^\infty \hat{f}_n e^{i\omega_n t} \Delta\! \,.
\end{equation}
Similarly, since $\omega_n$ becomes continuous as $T \rightarrow \infty$, we see that
\begin{equation}
    \hat{f}(\omega) \equiv \lim_{T \rightarrow \infty}  \hat{f}_n = \lim_{T \rightarrow \infty} \int_{-T}^T f(t)e^{-i \omega_n t}\, {\rm d}t. 
\end{equation}

\section{References}
\bibliography{references.bib}

\begin{thebibliography}{29}%
\makeatletter
\providecommand \@ifxundefined [1]{%
 \@ifx{#1\undefined}
}%
\providecommand \@ifnum [1]{%
 \ifnum #1\expandafter \@firstoftwo
 \else \expandafter \@secondoftwo
 \fi
}%
\providecommand \@ifx [1]{%
 \ifx #1\expandafter \@firstoftwo
 \else \expandafter \@secondoftwo
 \fi
}%
\providecommand \natexlab [1]{#1}%
\providecommand \enquote  [1]{``#1''}%
\providecommand \bibnamefont  [1]{#1}%
\providecommand \bibfnamefont [1]{#1}%
\providecommand \citenamefont [1]{#1}%
\providecommand \href@noop [0]{\@secondoftwo}%
\providecommand \href [0]{\begingroup \@sanitize@url \@href}%
\providecommand \@href[1]{\@@startlink{#1}\@@href}%
\providecommand \@@href[1]{\endgroup#1\@@endlink}%
\providecommand \@sanitize@url [0]{\catcode `\\12\catcode `\$12\catcode
  `\&12\catcode `\#12\catcode `\^12\catcode `\_12\catcode `\%12\relax}%
\providecommand \@@startlink[1]{}%
\providecommand \@@endlink[0]{}%
\providecommand \url  [0]{\begingroup\@sanitize@url \@url }%
\providecommand \@url [1]{\endgroup\@href {#1}{\urlprefix }}%
\providecommand \urlprefix  [0]{URL }%
\providecommand \Eprint [0]{\href }%
\providecommand \doibase [0]{http://dx.doi.org/}%
\providecommand \selectlanguage [0]{\@gobble}%
\providecommand \bibinfo  [0]{\@secondoftwo}%
\providecommand \bibfield  [0]{\@secondoftwo}%
\providecommand \translation [1]{[#1]}%
\providecommand \BibitemOpen [0]{}%
\providecommand \bibitemStop [0]{}%
\providecommand \bibitemNoStop [0]{.\EOS\space}%
\providecommand \EOS [0]{\spacefactor3000\relax}%
\providecommand \BibitemShut  [1]{\csname bibitem#1\endcsname}%
\let\auto@bib@innerbib\@empty
\bibitem [{\citenamefont {Sanderson}(2018)}]{3B1B}%
  \BibitemOpen
  \bibfield  {author} {\bibinfo {author} {\bibfnamefont {G.}~\bibnamefont
  {Sanderson}},\ }\href
  {https://www.3blue1brown.com/lessons/fourier-transforms} {\enquote {\bibinfo
  {title} {{3blue1brown.com}},}\ } (\bibinfo {year} {2018})\BibitemShut
  {NoStop}%
\bibitem [{\citenamefont {G{\"{u}}nther}(1973)}]{gunther-nmr}%
  \BibitemOpen
  \bibfield  {author} {\bibinfo {author} {\bibfnamefont {H.}~\bibnamefont
  {G{\"{u}}nther}},\ }\href@noop {} {\emph {\bibinfo {title} {{NMR
  Spectroscopy}}}},\ \bibinfo {edition} {3rd}\ ed.\ (\bibinfo  {publisher}
  {Wiley-VCH},\ \bibinfo {year} {1973})\BibitemShut {NoStop}%
\bibitem [{\citenamefont {Akitt}(1983)}]{borks-book}%
  \BibitemOpen
  \bibfield  {author} {\bibinfo {author} {\bibfnamefont {J.}~\bibnamefont
  {Akitt}},\ }\href {\doibase 10.1201/9781315274690} {\emph {\bibinfo {title}
  {{NMR and Chemistry}}}},\ \bibinfo {edition} {2nd}\ ed.\ (\bibinfo
  {publisher} {CRC Press},\ \bibinfo {address} {London},\ \bibinfo {year}
  {1983})\BibitemShut {NoStop}%
\bibitem [{\citenamefont {Griffiths}(1978)}]{transform-techniques}%
  \BibitemOpen
  \bibfield  {author} {\bibinfo {author} {\bibfnamefont {P.~R.}\ \bibnamefont
  {Griffiths}},\ }\href {\doibase 10.1007/978-1-4684-2403-4{\_}14} {\emph
  {\bibinfo {title} {{Transform Techniques in Chemistry}}}}\ (\bibinfo
  {publisher} {Springer New York},\ \bibinfo {year} {1978})\ pp.\ \bibinfo
  {pages} {355--378}\BibitemShut {NoStop}%
\bibitem [{\citenamefont {M{\"{u}}ller}(2021)}]{music}%
  \BibitemOpen
  \bibfield  {author} {\bibinfo {author} {\bibfnamefont {M.}~\bibnamefont
  {M{\"{u}}ller}},\ }\href {\doibase 10.1007/978-3-030-69808-9} {\emph
  {\bibinfo {title} {Fundamentals of Music Processing: Using Python and Jupyter
  Notebooks}}}\ (\bibinfo  {publisher} {Springer},\ \bibinfo {year} {2021})\
  pp.\ \bibinfo {pages} {1--495}\BibitemShut {NoStop}%
\bibitem [{\citenamefont {Brunton}\ and\ \citenamefont
  {Kutz}(2019)}]{brunton-book}%
  \BibitemOpen
  \bibfield  {author} {\bibinfo {author} {\bibfnamefont {S.~L.}\ \bibnamefont
  {Brunton}}\ and\ \bibinfo {author} {\bibfnamefont {J.~N.}\ \bibnamefont
  {Kutz}},\ }\href {\doibase 10.1017/9781108380690} {\emph {\bibinfo {title}
  {{Data-Driven Science and Engineering}}}}\ (\bibinfo  {publisher} {Cambridge
  University Press},\ \bibinfo {year} {2019})\BibitemShut {NoStop}%
\bibitem [{\citenamefont {Berger}(2002)}]{berger-mri}%
  \BibitemOpen
  \bibfield  {author} {\bibinfo {author} {\bibfnamefont {A.}~\bibnamefont
  {Berger}},\ }\bibfield  {title} {\enquote {\bibinfo {title} {{How does it
  work?: Magnetic resonance imaging}},}\ }\href {\doibase
  10.1136/bmj.324.7328.35} {\bibfield  {journal} {\bibinfo  {journal} {BMJ}\
  }\textbf {\bibinfo {volume} {324}},\ \bibinfo {pages} {35--35} (\bibinfo
  {year} {2002})}\BibitemShut {NoStop}%
\bibitem [{\citenamefont {Smith}(1985)}]{smith-mri}%
  \BibitemOpen
  \bibfield  {author} {\bibinfo {author} {\bibfnamefont {S.~L.}\ \bibnamefont
  {Smith}},\ }\bibfield  {title} {\enquote {\bibinfo {title} {{Nuclear Magnetic
  Resonance Imaging}},}\ }\href {\doibase 10.1021/ac00281a806} {\bibfield
  {journal} {\bibinfo  {journal} {Analytical Chemistry}\ }\textbf {\bibinfo
  {volume} {57}},\ \bibinfo {pages} {595A--608A} (\bibinfo {year}
  {1985})}\BibitemShut {NoStop}%
\bibitem [{\citenamefont {Oppenheim}\ and\ \citenamefont
  {Schafer}(2014)}]{discrete-time-signal-processing}%
  \BibitemOpen
  \bibfield  {author} {\bibinfo {author} {\bibfnamefont {A.~V.}\ \bibnamefont
  {Oppenheim}}\ and\ \bibinfo {author} {\bibfnamefont {R.~W.}\ \bibnamefont
  {Schafer}},\ }\href@noop {} {\emph {\bibinfo {title} {{Discrete-Time Signal
  Processing}}}},\ \bibinfo {edition} {3rd}\ ed.\ (\bibinfo  {publisher}
  {Pearson Education},\ \bibinfo {year} {2014})\BibitemShut {NoStop}%
\bibitem [{\citenamefont {Glasser}(1987{\natexlab{a}})}]{glasser-pt1}%
  \BibitemOpen
  \bibfield  {author} {\bibinfo {author} {\bibfnamefont {L.}~\bibnamefont
  {Glasser}},\ }\bibfield  {title} {\enquote {\bibinfo {title} {{Fourier
  Transforms for Chemists. Part I. Introduction to the Fourier Transform}},}\
  }\href {\doibase 10.1021/ed064pA228} {\bibfield  {journal} {\bibinfo
  {journal} {Journal of Chemical Education}\ }\textbf {\bibinfo {volume} {64}}
  (\bibinfo {year} {1987}{\natexlab{a}}),\ 10.1021/ed064pA228}\BibitemShut
  {NoStop}%
\bibitem [{\citenamefont {Glasser}(1987{\natexlab{b}})}]{glasser-pt2}%
  \BibitemOpen
  \bibfield  {author} {\bibinfo {author} {\bibfnamefont {L.}~\bibnamefont
  {Glasser}},\ }\bibfield  {title} {\enquote {\bibinfo {title} {{Fourier
  transforms for chemists. Part 2. Fourier transforms in chemistry and
  spectroscopy}},}\ }\href {\doibase 10.1021/ed064pA260} {\bibfield  {journal}
  {\bibinfo  {journal} {Journal of Chemical Education}\ }\textbf {\bibinfo
  {volume} {64}} (\bibinfo {year} {1987}{\natexlab{b}}),\
  10.1021/ed064pA260}\BibitemShut {NoStop}%
\bibitem [{\citenamefont {Glasser}(1987{\natexlab{c}})}]{glasser-pt3}%
  \BibitemOpen
  \bibfield  {author} {\bibinfo {author} {\bibfnamefont {L.}~\bibnamefont
  {Glasser}},\ }\bibfield  {title} {\enquote {\bibinfo {title} {{Fourier
  transforms for chemists. Part 3. Fourier transforms in data treatment}},}\
  }\href {\doibase 10.1021/ed064pA306} {\bibfield  {journal} {\bibinfo
  {journal} {Journal of Chemical Education}\ }\textbf {\bibinfo {volume} {64}}
  (\bibinfo {year} {1987}{\natexlab{c}}),\ 10.1021/ed064pA306}\BibitemShut
  {NoStop}%
\bibitem [{\citenamefont {Besal{\'{u}}}(2006)}]{graphical-rep}%
  \BibitemOpen
  \bibfield  {author} {\bibinfo {author} {\bibfnamefont {E.}~\bibnamefont
  {Besal{\'{u}}}},\ }\bibfield  {title} {\enquote {\bibinfo {title} {{A
  Graphical Presentation To Teach the Concept of the Fourier Transform}},}\
  }\href {\doibase 10.1021/ed083p1795} {\bibfield  {journal} {\bibinfo
  {journal} {Journal of Chemical Education}\ }\textbf {\bibinfo {volume}
  {83}},\ \bibinfo {pages} {1795--1797} (\bibinfo {year} {2006})}\BibitemShut
  {NoStop}%
\bibitem [{\citenamefont {Baiz}\ \emph {et~al.}(2024)\citenamefont {Baiz},
  \citenamefont {Berger}, \citenamefont {Donald}, \citenamefont {de~Paula},
  \citenamefont {Fried}, \citenamefont {Rubenstein}, \citenamefont {Stokes},
  \citenamefont {Takematsu},\ and\ \citenamefont {Londergan}}]{LABSIP}%
  \BibitemOpen
  \bibfield  {author} {\bibinfo {author} {\bibfnamefont {C.~R.}\ \bibnamefont
  {Baiz}}, \bibinfo {author} {\bibfnamefont {R.~F.}\ \bibnamefont {Berger}},
  \bibinfo {author} {\bibfnamefont {K.~J.}\ \bibnamefont {Donald}}, \bibinfo
  {author} {\bibfnamefont {J.~C.}\ \bibnamefont {de~Paula}}, \bibinfo {author}
  {\bibfnamefont {S.~D.}\ \bibnamefont {Fried}}, \bibinfo {author}
  {\bibfnamefont {B.}~\bibnamefont {Rubenstein}}, \bibinfo {author}
  {\bibfnamefont {G.~Y.}\ \bibnamefont {Stokes}}, \bibinfo {author}
  {\bibfnamefont {K.}~\bibnamefont {Takematsu}}, \ and\ \bibinfo {author}
  {\bibfnamefont {C.}~\bibnamefont {Londergan}},\ }\bibfield  {title} {\enquote
  {\bibinfo {title} {{Lowering Activation Barriers to Success in Physical
  Chemistry (LABSIP): A Community Project}},}\ }\href {\doibase
  10.1021/acs.jpca.3c07015} {\bibfield  {journal} {\bibinfo  {journal} {Journal
  of Physical Chemistry A}\ }\textbf {\bibinfo {volume} {128}},\ \bibinfo
  {pages} {3--9} (\bibinfo {year} {2024})}\BibitemShut {NoStop}%
\bibitem [{\citenamefont {{Mark Newman}}(2013)}]{comp-physics-newman}%
  \BibitemOpen
  \bibfield  {author} {\bibinfo {author} {\bibnamefont {{Mark Newman}}},\
  }\href@noop {} {\emph {\bibinfo {title} {{Computational Physics}}}},\
  \bibinfo {edition} {revised and expanded}\ ed.\ (\bibinfo  {publisher}
  {CreateSpace Independent Publishing Platform},\ \bibinfo {address}
  {Washington},\ \bibinfo {year} {2013})\ pp.\ \bibinfo {pages}
  {289--326}\BibitemShut {NoStop}%
\bibitem [{\citenamefont {Pres}\ \emph {et~al.}(1992)\citenamefont {Pres},
  \citenamefont {Teukolsky}, \citenamefont {Vetterling},\ and\ \citenamefont
  {Flannery}}]{num-recipes}%
  \BibitemOpen
  \bibfield  {author} {\bibinfo {author} {\bibfnamefont {W.~H.}\ \bibnamefont
  {Pres}}, \bibinfo {author} {\bibfnamefont {S.~A.}\ \bibnamefont {Teukolsky}},
  \bibinfo {author} {\bibfnamefont {W.~T.}\ \bibnamefont {Vetterling}}, \ and\
  \bibinfo {author} {\bibfnamefont {B.~P.}\ \bibnamefont {Flannery}},\
  }\href@noop {} {\emph {\bibinfo {title} {{Numerical Recipes in C}}}},\
  \bibinfo {edition} {2nd}\ ed.\ (\bibinfo  {publisher} {Cambridge University
  Press},\ \bibinfo {year} {1992})\ p.\ \bibinfo {pages} {496}\BibitemShut
  {NoStop}%
\bibitem [{\citenamefont {Griffiths}\ and\ \citenamefont
  {De~Haseth}(2007)}]{IR-book}%
  \BibitemOpen
  \bibfield  {author} {\bibinfo {author} {\bibfnamefont {P.~R.}\ \bibnamefont
  {Griffiths}}\ and\ \bibinfo {author} {\bibfnamefont {J.~A.}\ \bibnamefont
  {De~Haseth}},\ }\href {\doibase 10.1002/047010631X} {\emph {\bibinfo {title}
  {{Fourier Transform Infrared Spectrometry}}}},\ \bibinfo {edition} {2nd}\
  ed.\ (\bibinfo  {publisher} {Wiley-Interscience},\ \bibinfo {year}
  {2007})\BibitemShut {NoStop}%
\bibitem [{\citenamefont {Boas}(2005)}]{boas}%
  \BibitemOpen
  \bibfield  {author} {\bibinfo {author} {\bibfnamefont {M.~L.}\ \bibnamefont
  {Boas}},\ }\href@noop {} {\emph {\bibinfo {title} {{Mathematical Methods in
  the Physical Sciences}}}},\ \bibinfo {edition} {3rd}\ ed.\ (\bibinfo
  {publisher} {Wiley},\ \bibinfo {year} {2005})\BibitemShut {NoStop}%
\bibitem [{\citenamefont {Tolstov}(1976)}]{fourierseries-georgi}%
  \BibitemOpen
  \bibfield  {author} {\bibinfo {author} {\bibfnamefont {G.~P.}\ \bibnamefont
  {Tolstov}},\ }\href@noop {} {\emph {\bibinfo {title} {{Fourier Series}}}}\
  (\bibinfo  {publisher} {Dover Publications},\ \bibinfo {year}
  {1976})\BibitemShut {NoStop}%
\bibitem [{\citenamefont {Stewart}(2020)}]{stewart}%
  \BibitemOpen
  \bibfield  {author} {\bibinfo {author} {\bibfnamefont {J.}~\bibnamefont
  {Stewart}},\ }\href@noop {} {\emph {\bibinfo {title} {{Calculus: Early
  Transcendentals}}}},\ \bibinfo {edition} {9th}\ ed.\ (\bibinfo  {publisher}
  {Cengage Learning},\ \bibinfo {year} {2020})\BibitemShut {NoStop}%
\bibitem [{\citenamefont {Pons}(2014)}]{pons}%
  \BibitemOpen
  \bibfield  {author} {\bibinfo {author} {\bibfnamefont {M.~A.}\ \bibnamefont
  {Pons}},\ }\href {\doibase 10.1007/978-1-4614-9638-0} {\emph {\bibinfo
  {title} {{Real Analysis for the Undergraduate}}}}\ (\bibinfo  {publisher}
  {Springer New York},\ \bibinfo {address} {New York, NY},\ \bibinfo {year}
  {2014})\BibitemShut {NoStop}%
\bibitem [{Note1()}]{Note1}%
  \BibitemOpen
  \bibinfo {note} {For completeness, we note that the above integrals exist
  provided that $\protect \qopname \relax m{lim}_{t \rightarrow \pm \infty
  }f(t) = 0$ and $\protect \qopname \relax m{lim}_{\omega \rightarrow \pm
  \infty }\protect \hat {f}(\omega ) = 0$, respectively, and $f$ and $\protect
  \hat {f}$ are absolutely integrable (i.e., $\DOTSI \intop \ilimits@ _{-\infty
  }^\infty |f(t)| \protect \, {\protect \rm d}t < \infty $ and $\DOTSI \intop
  \ilimits@ _{-\infty }^\infty |\protect \hat {f}(\omega )| \protect \,
  {\protect \rm d}\omega < \infty $). Generally, spectroscopic measurements
  (e.g., FIDs, interferograms) meet these requirements.}\BibitemShut {Stop}%
\bibitem [{\citenamefont {Strang}(1994)}]{strang-quote}%
  \BibitemOpen
  \bibfield  {author} {\bibinfo {author} {\bibfnamefont {G.}~\bibnamefont
  {Strang}},\ }\bibfield  {title} {\enquote {\bibinfo {title} {{Wavelets}},}\
  }\href@noop {} {\bibfield  {journal} {\bibinfo  {journal} {American
  Scientist}\ }\textbf {\bibinfo {volume} {82}},\ \bibinfo {pages} {250--255}
  (\bibinfo {year} {1994})}\BibitemShut {NoStop}%
\bibitem [{\citenamefont {Stone}(2005)}]{moments}%
  \BibitemOpen
  \bibfield  {author} {\bibinfo {author} {\bibfnamefont {N.~J.}\ \bibnamefont
  {Stone}},\ }\bibfield  {title} {\enquote {\bibinfo {title} {{Table of nuclear
  magnetic dipole and electric quadrupole moments}},}\ }\href {\doibase
  10.1016/j.adt.2005.04.001} {\bibfield  {journal} {\bibinfo  {journal} {Atomic
  Data and Nuclear Data Tables}\ }\textbf {\bibinfo {volume} {90}},\ \bibinfo
  {pages} {75--176} (\bibinfo {year} {2005})}\BibitemShut {NoStop}%
\bibitem [{\citenamefont {Katai}\ and\ \citenamefont
  {Toth}(2010)}]{multisensorylearning}%
  \BibitemOpen
  \bibfield  {author} {\bibinfo {author} {\bibfnamefont {Z.}~\bibnamefont
  {Katai}}\ and\ \bibinfo {author} {\bibfnamefont {L.}~\bibnamefont {Toth}},\
  }\bibfield  {title} {\enquote {\bibinfo {title} {{Technologically and
  artistically enhanced multi-sensory computer-programming education}},}\
  }\href {\doibase 10.1016/j.tate.2009.04.012} {\bibfield  {journal} {\bibinfo
  {journal} {Teaching and Teacher Education}\ }\textbf {\bibinfo {volume}
  {26}},\ \bibinfo {pages} {244--251} (\bibinfo {year} {2010})}\BibitemShut
  {NoStop}%
\bibitem [{chi(2016)}]{chirp}%
  \BibitemOpen
  \href@noop {} {\enquote {\bibinfo {title}
  {{https://betterworld.mit.edu/spectrum/issues/spring-2016/the-chirp-heard-across-the-universe/}},}\
  } (\bibinfo {year} {2016})\BibitemShut {NoStop}%
\bibitem [{\citenamefont {{NIST Chemistry WebBook}}(1977)}]{3pentanone}%
  \BibitemOpen
  \bibfield  {author} {\bibinfo {author} {\bibnamefont {{NIST Chemistry
  WebBook}}},\ }\href
  {https://webbook.nist.gov/cgi/inchi?ID=C96220&SPEC&Index=1#IR-SPEC} {\enquote
  {\bibinfo {title} {{https://webbook.nist.gov/}},}\ } (\bibinfo {year}
  {1977})\BibitemShut {NoStop}%
\bibitem [{\citenamefont {Prabhu}(2018)}]{windowing}%
  \BibitemOpen
  \bibfield  {author} {\bibinfo {author} {\bibfnamefont {K.~M.~M.}\
  \bibnamefont {Prabhu}},\ }\href {\doibase 10.1201/9781315216386} {\emph
  {\bibinfo {title} {{Window Functions and Their Applications in Signal
  Processing}}}},\ \bibinfo {edition} {1st}\ ed.\ (\bibinfo  {publisher} {CRC
  Press},\ \bibinfo {address} {Boca Raton},\ \bibinfo {year}
  {2018})\BibitemShut {NoStop}%
\bibitem [{\citenamefont {{Mestrelab Research}}(2024)}]{mnova}%
  \BibitemOpen
  \bibfield  {author} {\bibinfo {author} {\bibnamefont {{Mestrelab
  Research}}},\ }\href {https://mestrelab.com/} {\enquote {\bibinfo {title}
  {{https://mestrelab.com/}},}\ } (\bibinfo {year} {2024})\BibitemShut
  {NoStop}%
\end{thebibliography}%

\end{document}